\documentclass[12pt]{article}
\usepackage{graphicx}
\usepackage{wrapfig}
\usepackage{lipsum}  
\usepackage{accents}
\usepackage{subfigure}

\newcommand\pubnumber{NuPhys2016-Tsai}
\newcommand\pubdate{\today}
\newcommand\brabar{\scalebox{.3}{(}\raisebox{-1.7pt}{$-$}\scalebox{.3}{)}}

\def\slac{SLAC National Accelerator Laboratory, Menlo Park, CA, USA}

\def\Title#1{\begin{center} {\Large #1 } \end{center}}
\def\Author#1{\begin{center}{ \sc #1} \end{center}}
\def\Address#1{\begin{center}{ \it #1} \end{center}}

\newcommand\pubblock{\rightline{\begin{tabular}{l} \pubnumber\\
         \pubdate  \end{tabular}}}
\newenvironment{Abstract}{\begin{quotation}  }{\end{quotation}}
\newenvironment{Presented}{\begin{quotation} \begin{center} 
             PRESENTED AT\end{center}\bigskip 
      \begin{center}\begin{large}}{\end{large}\end{center} \end{quotation}}

\def\beq{\begin{equation}}
\def\eeq#1{\label{#1}\end{equation}}
\def\eeqn{\end{equation}}

\def\beqa{\begin{eqnarray}}
\def\eeqa#1{\label{#1}\end{eqnarray}}
\def\eeqan{\end{eqnarray}}

\let\bar=\overbar

\def\Dslash{\not{\hbox{\kern-4pt $D$}}}
\def\dslash{\not{\hbox{\kern-2pt $\del$}}}

\def\msb{{\bar{\ssstyle M \kern -1pt S}}}

\begin{document}
\begin{titlepage}
\pubblock

\vfill
\Title{MicroBooNE and its Cross Section Measurement}
\vfill
\Author{Yun-Tse Tsai, for the MicroBooNE collaboration}
\Address{\slac}
\vfill
\begin{Abstract}
MicroBooNE (the Micro Booster Neutrino Experiment) is a short-baseline
neutrino experiment 
based on the technology of a liquid-argon time-projection chamber (LArTPC),
and has recently completed its first year of data-taking in the Fermilab
Booster Neutrino Beam.
It aims to address the anomalous excess of events with an electromagnetic final
state in MiniBooNE, to measure neutrino-argon interaction cross sections,
and to provide relevant R\&D for the future LArTPC experiments, such as DUNE.
In these proceedings, we present the first reconstructed energy spectrum of Michel
electrons from cosmic muon decays, the first kinematic distributions of the candidate
muon tracks from $\nu_{\mu}$-argon charged-current interactions, and a
demonstration of an electromagnetic shower reconstruction from $\pi^0$s produced
by $\nu_{\mu}$-argon charged-current interactions.
The results demonstrate the first fully automated reconstruction and selection
algorithms in a large LArTPC and serve as foundations for future measurements.

\end{Abstract}
\vfill
\begin{Presented}
NuPhys2016, Prospects in Neutrino Physics\\
Barbican Centre, London, UK,  December 12--14, 2016
\end{Presented}
\vfill
\end{titlepage}
\def\thefootnote{\fnsymbol{footnote}}
\setcounter{footnote}{0}

\section{The MicroBooNE Experiment}

The MicroBooNE experiment is a neutrino experiment aiming to measure
oscillation of neutrino flavors and neutrino-nuclear interaction cross
sections.
Located in the Booster Neutrino Beam (BNB) at Fermilab at a baseline of
470~m, MicroBooNE is the first experiment in the U.S. utilizing a
large liquid-argon time-projection chamber (LArTPC)~\cite{uboone_tdr}.\\

The primary physics goal of MicroBooNE is to address the excess of data 
events with an electromagnetic object in the regime of neutrino energy of
200 -- 500~MeV reported by the MiniBooNE experiment~\cite{mboone_lee}.
As a Cherenkov detector filled with mineral oil, MiniBooNE was not able
to distinguish electrons from photons.
If the excess comes from events with an electron,
it may imply existence of a sterile neutrino from the interpretation of
the $\accentset{\brabar}{\nu_{\mu}}\to\accentset{\brabar}{\nu_e}$ oscillation. 
On the other hand, if the electromagnetic object in those events is a photon, 
it may indicate an unknown background component.
A LArTPC is able to distinguish electrons and photons by looking for 
the $\gamma\to e^+e^-$ topology at the start of an electromagnetic shower,
and can thereby be exploited to investigate the MiniBooNE anomaly.\\

Measuring neutrino-argon interaction cross sections is another goal
of the MicroBooNE experiment.
Neutrino-nuclear interactions are currently not well understood,
and have significant impacts on the precision of
neutrino oscillation measurements.
In particular, one of the most important neutrino experiments in the next
generation, Deep Underground Neutrino Experiment (DUNE)~\cite{DUNE}, aims to address
the CP-invariance violation in the lepton sector and the neutrino mass
ordering by measuring rates and energy spectra of
neutrino oscillation with the LArTPC technology.  Therefore, precise
measurements of neutrino-argon cross sections at MicroBooNE will be relevant.\\

In addition, the MicroBooNE detector is utilized to explore astroparticle
and exotic physics, such as the detection of neutrinos from core-collapse supernova 
explosions, and searches for nucleon decays, nucleon oscillation, as well as
dark matter candidates.
The size of the MicroBooNE detector is not sensitive to most of
the searches. 
However, we can demonstrate the technique, study backgrounds, and
probe the thresholds of large LArTPCs.
Moreover, LArTPC R\&D and detector physics can be performed with MicroBooNE.
For example, we demonstrate the purification of liquid argon, the design of the high
voltage system, the cold electronics, readout and data acquisition systems.
The effects of detector noise, electron recombination and attenuation can also be
characterized at MicroBooNE.
All the results will provide the DUNE experiment
with valuable information.

\section{Detector and its Performance}

\begin{figure}[htb]
\centering
\subfigure[]{
\centering
\includegraphics[width=0.45\textwidth]{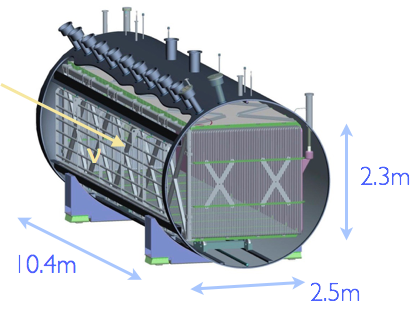}
\label{fig:MicroBooNE}
}
\hspace{0.5cm}
\subfigure[]{
\centering
\includegraphics[width=0.45\textwidth]{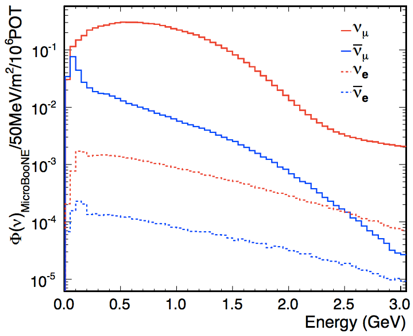}
\label{fig:flux}
}
\caption{(a) Schematic drawing of the MicroBooNE cryostat which hosts the TPC.
(b) The expected components and energy spectra of the BNB flux at MicroBooNE 
in the neutrino mode.}
\end{figure}

MicroBooNE consists of one time-projection chamber (TPC), with
89~tons of liquid argon in its active volume.
As shown in Fig.~\ref{fig:MicroBooNE},
the TPC has dimensions of 10.4~m in the BNB direction, 2.3~m in vertical,
and 2.5~m between the cathode and anode, along which a high voltage electric field
of 273~V/cm is applied.
There are three wire planes at the anode, each oriented by a degree of $\pm60^{\circ}$
and $0^{\circ}$ with respect to the vertical, reading out the deposited
charges at 2~MHz.
A light collection system consisting of 32 8-inch photomultiplier tubes
(PMTs) is mounted behind the anode wire planes.
More details about the detector can be found in~\cite{uboone_detector}.\\

\begin{figure}[htb]
\centering
\includegraphics[width=0.9\linewidth]{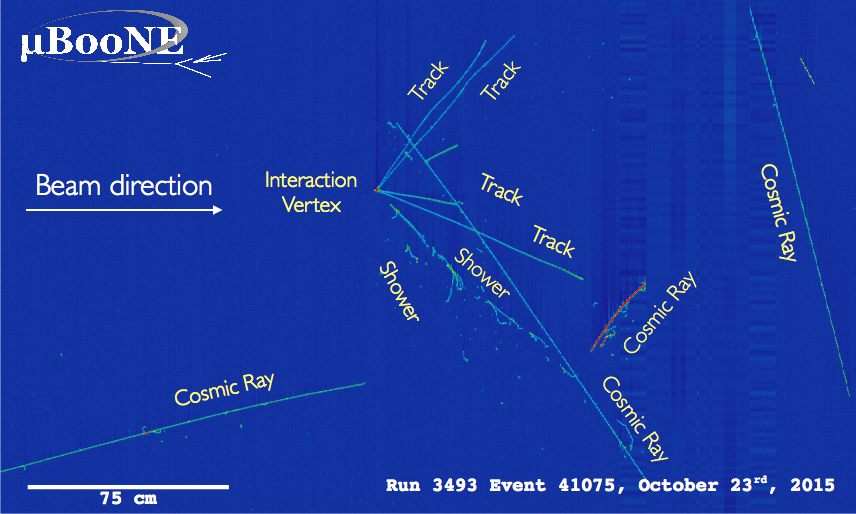}
\caption{An annotated display of an event from MicroBooNE data, taken
from the third wire plane, where the wires are vertical.
The color scale indicates the amount of deposited charges.
The horizontal direction of the display represents the wire numbers,
while the vertical direction shows the drift time, or the drift distance.
The display characterizes a neutrino-argon interaction with a few tracks
and two electromagnetic showers in the final state, overlaid with multiple
tracks from cosmic rays.}
\label{fig:uboone_evd}
\end{figure}

Charged particles produced from neutrino-argon interactions ionize
argon atoms and create scintillation light.
The scintillation light, produced in a time scale of 6~nanoseconds,
is collected by the PMTs, determining the event timing,
while the ionization electrons slowly drift towards the anode.
Those electrons pass through the first two wire planes, leaving
induced current, and then are collected in the third wire planes.
Assuming a constant drift velocity ($\sim 1.1$~mm/$\mu$s), 
the electron drift time is proportional to
the drift distance, and therefore each wire plane provides a two-dimensional
image with the granularity of 3~mm (wire pitch) times 0.6~mm (the sampling rate
of digitization under the current high voltage configuration).
Fig.~\ref{fig:uboone_evd} illustrates the high spatial resolution of MicroBooNE
LArTPC and the capability of characterizing a complicated event topology.
A three-dimensional event can be reconstructed from the three two-dimensional
images from the three wire planes.\\

\begin{figure}[htb]
\centering
\includegraphics[width=0.8\linewidth]{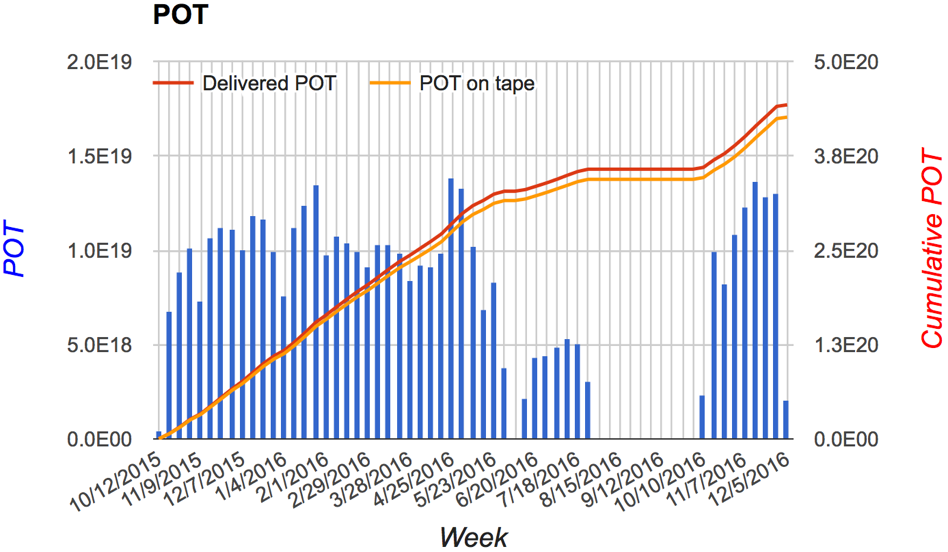}
\caption{The recorded protons on target (POT) per week (histograms) and
the cumulatively delivered and recorded POT (curves).}
\label{fig:POT}
\end{figure}

MicroBooNE started taking BNB neutrino data on October 15th, 2015.
The composition of the BNB flux, dominated by $\nu_{\mu}$,
can be found in Fig.~\ref{fig:flux}.
The detector and the data acquisition have performed 
stably~\cite{uboone_stability}, and 
$4\times 10^{20}$ protons on target (POT) have been recorded
as of the date of the conference.
Fig.~\ref{fig:POT} shows the cumulative efficiency of the data acquisition
at MicroBooNE.

\section{Reconstruction of Physics Objects}
\label{sec:reco}

Fully automated reconstruction algorithms are required to tackle the 
great amount of charge deposition from particles produced in neutrino-argon 
interactions, and from particles induced by cosmic rays during the
long readout window (4.8~milliseconds) in an event.
They are also needed to reduce the bias introduced by a visual scan.
To process data collected at the TPC,
we start with filtering the noise from the detector electronics~\cite{uboone_noise}, 
and then extract hits from the digital waveforms.
Multiple clustering algorithms are applied, associating the hits 
originating from the same charged particles~\cite{uboone_pandora}.\\

\begin{figure}[htb]
\subfigure[]{
\centering
\includegraphics[width=0.6\linewidth]{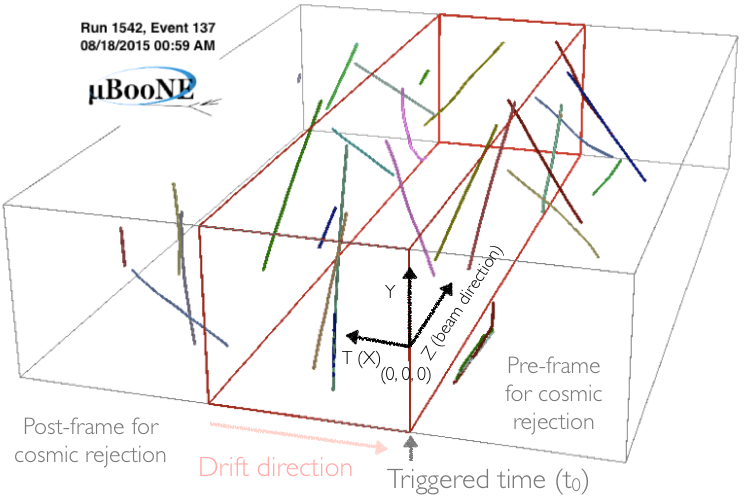}
\label{fig:TrackReco}}
\hspace{0.5cm}
\subfigure[]{
\centering
\includegraphics[width=0.35\linewidth]{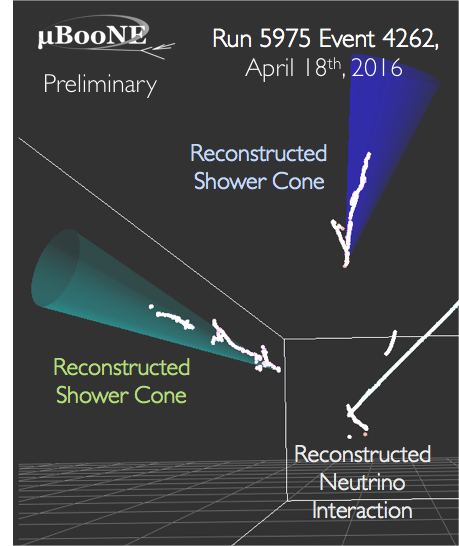}
\label{fig:ShowerReco}}
\caption{Three-dimensional reconstructed events from data collected at 
MicroBooNE: (a) An event containing cosmic rays collected outside the BNB window.
The three boxes show the full readout length per event, corresponding to 
4.8~milliseconds.  The red highlighted box outlines the 1.6~milliseconds
after the trigger time.
The colored lines represent reconstructed tracks from cosmic rays.
(b) An event containing a $\nu_{\mu}$-argon charged-current interaction
with a $\pi^0$ production, selected by a visual scan.  
The white points are the reconstructed locations
of deposited charges in the three-dimensional volume in the TPC.
The colored cones represent the geometry of the reconstructed electromagnetic
showers, possibly originating from the photons from the decay of the $\pi^0$.
}
\end{figure}

We use a three-dimensional track fitter to reconstruct tracks and remove
hits associated to through-going tracks, which likely represent cosmic rays.
Fig.~\ref{fig:TrackReco} illustrates reconstructed tracks in
an event taken outside the BNB operation window.
Subsequently, we reconstruct the remaining hits and obtain tracks, electromagnetic
showers, and neutrino-argon interaction vertices.
An event containing a neutrino-argon interaction with reconstructed tracks 
and showers in the
final state can be found in Fig.~\ref{fig:ShowerReco}.

\section{First Analyses}

Utilizing the reconstructed physics objects, we develop different selection criteria
for various analyses.
In this section, the first analyses from MicroBooNE will be discussed.

\subsection{Michel Electrons}

To further understand the detector response in the tens of MeV energy range
and the muon identification,
we study the energy spectrum of Michel electrons, electrons in the decay
products of stopping muons~\cite{uboone_Michel}.\\

In this analysis, we use muons from cosmic rays.
The data sample contains 280,751 events collected outside of the BNB operation
windows, corresponding to 1,347 seconds.
A set of clustering algorithms are developed, profiling the deposited
charges based on the highly resoluted topological and calorimetric information
provided by the third wire plane, which collects ionization electrons.
We identify the tracks as stopping muons by looking for an increase in the 
charge deposition per unit length towards the end of the track, 
and the electron candidate is recognized as
the track coming after the identified
muon stopping point at an angle with respect to the muon track.
The reconstruction and selection algorithms are fully automated.\\

The energy of Michel electrons is calculated from the reconstructed charges
at the third (collection) wire plane with 
appropriate electronic calibration factors and correction
factors.
The correction factors account for two effects,
\begin{itemize}
\item recombination of argon ions and ionization electrons,
\item attenuation of the ionization electrons during the drift path owing
to the electronegative contamination in the liquid argon.
\end{itemize}
In this analysis, constant correction factors are applied;
further development and studies are underway to better model the two
effects and the corresponding correction factors.\\

\begin{figure}[htb]
\centering
\includegraphics[width=0.5\linewidth]{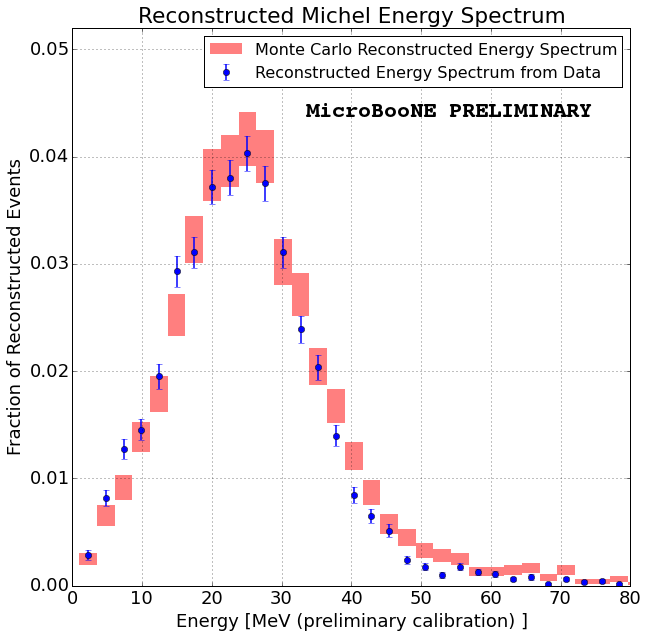}
\caption{The reconstructed energy spectrum of Michel electrons from data and Monte
Carlo simulation.
The uncertainty in both data and Monte Carlo simulation accounts for the statistic
uncertainty.}
\label{fig:Michel}
\end{figure}

The distribution of the reconstructed energy from Michel electrons is shown
in Fig.~\ref{fig:Michel}.
The energy distribution of Michel electrons typically has a sharp edge at 
52.2~MeV, half the mass of muons.
The distortion of the spectrum is owing to the fact that the radiated photons,
which can start the $e^+e^-$ pair production in tens of centimeters,
are poorly included in the energy reconstruction.
Nonetheless, the reasonable agreement between the spectra from the Monte Carlo (MC)
simulation and the data demonstrates our understanding of Michel electrons in
LArTPCs.
The remaining difference in the two spectra may come from variation of calibration
factors in different TPC wire channels.
Further studies are ongoing to improve the analysis.

\subsection{$\nu_{\mu}$-argon Charged-current Interactions}

The measurement of the $\nu_{\mu}$-argon charged-current interaction cross section
provides us with a foundation for comparisons to theoretical calculations
and other experimental results.
In addition, it serves as a common starting point for further measurements of exclusive
interaction channels, such as the charged-current interaction with the
production of a $\pi^0$.
Owing to the surface location, the recorded events at MicroBooNE are dominated by
cosmic rays, and identifying and removing those background events are challenging.
In this analysis, we present multiple kinematic distributions of muons produced
from the $\nu_{\mu}$-argon charged-current interaction.
The analysis outlines required tools for data quality and detector stability checks,
for physics object reconstructions and event selections.
It also guides us towards strategies and required improvements for all
MicroBooNE analyses~\cite{uboone_CCincl}.\\

A data sample of 546,910 events, corresponding to $4.95\times10^{19}$
protons on target, is analyzed.
The MC event generator GENIE is used to simulate the neutrino-argon 
interactions, while the particles induced by cosmic rays in these events are
modeled by the CORSIKA simulation program.
The passages of particles through the detector are simulated by GEANT4. 
Further, we exploit data collected outside the BNB operation windows for
an estimation of the background events containing no neutrino-argon interaction.\\

Utilizing the reconstruction algorithms described in Sec.~\ref{sec:reco},
we develop fully automated selection schemes.
As a $\nu_{\mu}$-argon charged-current interaction produces a muon, leaving
a long track in the detector, we select such events coincident with the BNB beam
timing.
One of the schemes and the consequent kinematic distributions are presented
in these proceedings.
We require
\begin{enumerate}
\item a light signal above 50 photoelectrons (P.E.) within the BNB 
      operation window (1.6~$\mu$s, much shorter than the TPC readout window, 4.8~ms), 
      indicating activities coincident with the beam timing,
\item at least a track longer than 70~cm, identifying as the muon candidate,
\item a light signal above 50~P.E. in agreement with the position of the candidate
      muon track in the beam direction,
\item the reconstructed interaction vertex within the fiducial volume (20~cm
      from the border in vertical, and 10~cm from the border in the other 
      dimensions), removing events induced by cosmic rays or other background
      interactions,
\item at least a track starting within 3~cm from the interaction vertex,
      ensuring the production of charged particles near the vertex,
\item multiple sets of selection criteria on track kinematics for different
      charged particle multiplicities.
\end{enumerate}
More details can be found in~\cite{uboone_CCincl}.\\

\begin{figure}[h]
\centering
\subfigure[]{
\centering
\includegraphics[width=0.45\linewidth]{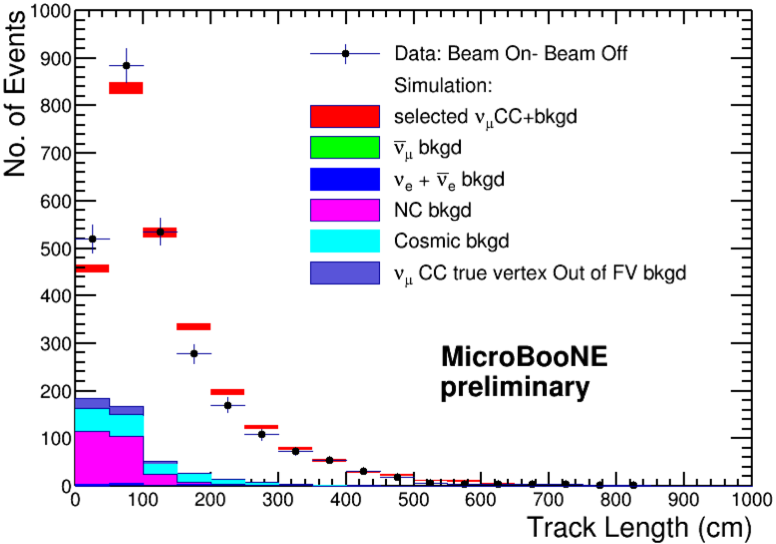}} \\
\subfigure[]{
\centering
\includegraphics[width=0.45\linewidth]{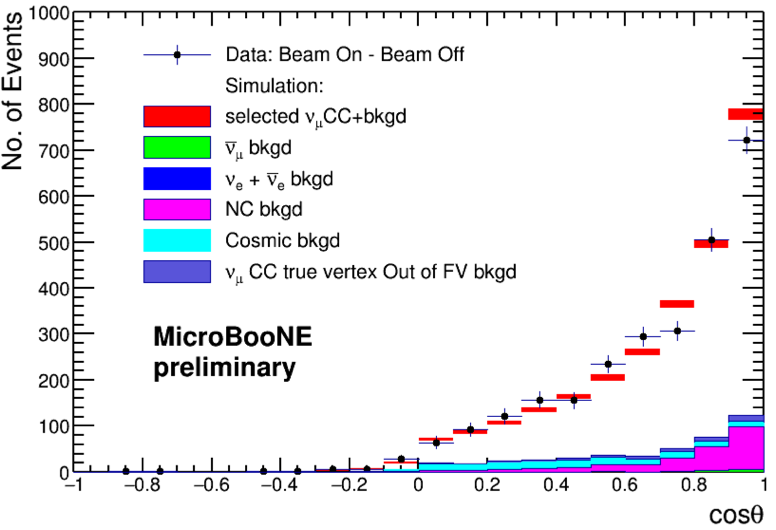}}
\subfigure[]{
\centering
\includegraphics[width=0.45\linewidth]{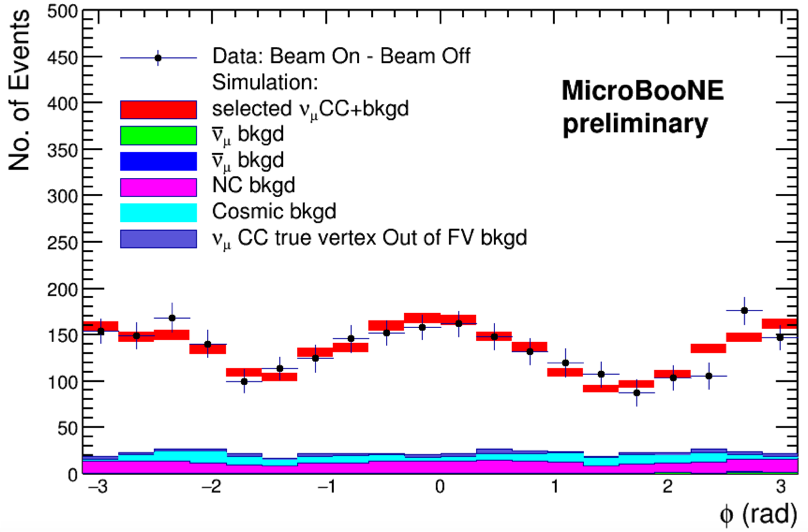}}
\caption{Kinematic distributions of the candidate muon tracks in the selected
events: (a) track length, (b) $\cos\theta$, where $\theta$ denotes the angle
of the track with respect to the neutrino beam direction, and (c) $\phi$,
the azimuthal angle around the beam direction.
The number of events in the simulation is normalized to that in the data.
Events from pure cosmic rays are subtracted.
The uncertainty in the Monte Carlo simulation and the data accounts for the
statistical uncertainty only.
The pions produced from neutral-current interactions misidentified as
muon tracks are typically shorter, and contribute to the first two bins in
the track length distribution.
The efficiency at $\phi = \pm \pi/2$, corresponding to the vertical direction,
is lower because the candidate muon tracks in vertical are more likely to
be identified as cosmic rays and are removed.}
\label{fig:CCIncl}
\end{figure}

In the selected sample, we obtain
an efficiency convoluted with acceptance of 30\%, while obtaining
the purity of 65\%.
The dominant background events originate from cosmic rays.
Fig.~\ref{fig:CCIncl} shows kinematic distributions of the candidate
muon tracks, including the length, the angle with respect to the neutrino
beam direction ($\theta$), and the azimuthal angle around the beam direction
($\phi$).
The pure cosmic ray background events, determined using the data collected outside of
the BNB operation window, have been subtracted from those distributions.\\

The distributions from the MC simulation agree reasonably well with those from the data,
indicating our capability of modeling the signal, background events,
as well as the detector response.
Studies of systematic uncertainties in MicroBooNE are currently underway;
nonetheless, we expect the major contributions of the systematic uncertainties
would originate from the modeling of the BNB flux, the detector effects
(e.g. the detector noise, non-uniformity of the electric field),
and the simulation of neutrino-nucleus interactions.

\subsection{$\nu_{\mu}$-argon Charged-current Interactions with $\pi^0$ Production}

\begin{figure}[htb]
\subfigure[]{
\centering
\includegraphics[width=0.45\linewidth]{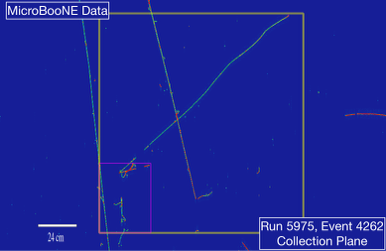}
\label{fig:CCpi0EVD}}
\hspace{0.5cm}
\subfigure[]{
\centering
\includegraphics[width=0.45\linewidth]{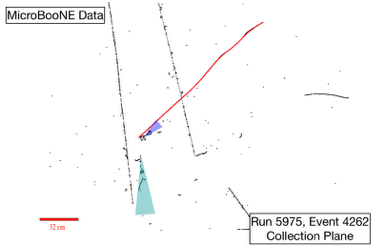}
\label{fig:CCpi0Reco}}
\caption{An event containing a $\nu_{\mu}$-argon charged-current interaction
with a $\pi^0$ production candidate, selected by a visual scan: (a) display with raw 
digital waveforms collected at the third wire plane, 
(b) display with reconstructed hits (black points), tracks (red lines), 
and electromagnetic showers (colored triangles) projected into the third
wire plane.}
\label{fig:CCpi0}
\end{figure}

Reconstruction algorithms for electromagnetic showers are the key step towards
the $\nu_{\mu}\to\nu_e$ oscillation analysis.
The reconstructed invariant $\pi^0$ mass is important to 
demonstrate the performance of our electromagnetic shower reconstruction algorithms,
as it requires both the direction and the energy of the reconstructed electromagnetic
showers originating from the photons from the decays of $\pi^0$s.
Out of the candidate events of $\nu_{\mu}$-argon charged-current
interactions, we visually select a few events potentially containing a $\pi^0$ production.
As illustrated in Fig.~\ref{fig:CCpi0EVD}, the $\pi^0$ decays into two photons,
each travels a few to a few tens of centimeters and then starts developing an electromagnetic
shower.
We thereby identify events containing two detached electromagnetic showers pointing 
back to the interaction vertex, which can be anchored by the beginning of the
candidate muon track~\cite{uboone_CCpi0}.\\

Fig.~\ref{fig:CCpi0Reco} shows the reconstructed shower cones and the hits
used to form the cones.
Developments on calorimetry and studies on systematic uncertainties are currently
underway.
Further, we have made progress towards fully automated selection exclusively on
this interaction channel with a fair efficiency, and plan to obtain results in
the near future.

\section{Outlook and Summary}

MicroBooNE is unique in its physics goals of addressing the anomaly reported
by the MiniBooNE experiment, of measuring neutrino-argon cross sections,
and of conducting R\&D for both astroparticle and exotic physics searches and
LArTPC performances.
It has been fully operational and stably taking neutrino data for 10 months, 
recording $4.2\times 10^{20}$ POT on tape.
In these proceedings, we present the first fully automated reconstruction
and event selection algorithms for LArTPCs and the first results, 
the energy spectrum of
Michel electrons, the kinematic distributions of $\nu_{\mu}$-argon 
charged-current interactions, and the demonstration of the shower reconstruction
algorithms with the events containing a $\pi^0$ production.
The energy spectrum of Michel electrons is a standard tool
used for energy calibrations, while the distributions and the demonstration
of the $\nu_{\mu}$-argon charged-current interactions guide us towards the
developments and studies for the final cross section measurements and
the neutrino oscillation analyses.\\

It is important to precisely measure the neutrino-nucleus cross sections
as their uncertainties are the major
components of the systematic uncertainties of the neutrino oscillation
measurements.
As of today, only few neutrino-argon cross sections have been
reported~\cite{pdg_nuxsec}, 
and therefore the cross sections measured by MicroBooNE would
significantly improve our knowledge.
In particular, the energy regime of neutrinos produced by BNB, from 200~MeV
to 2~GeV, corresponds to the second oscillation maximum of the DUNE
experiment.
We plan to deliver an inclusive neutrino-argon cross section measurement
in 2017, and will obtain several measurements of exclusive and differential cross 
sections.\\


In 2018/2019, the Short Baseline Neutrino (SBN) Program will be
operational, aiming to answer the question of existence of sterile neutrinos,
which could potentially explain the MiniBooNE anomaly in the low energy regime and 
the earlier anomaly
reported by the Liquid Scintillator Neutrino Detector (LSND)~\cite{SBN}.
The SBN program will utilize the BNB neutrino beam as the neutrino source,
and will contain a near and a far detectors to MicroBooNE.
The Short Baseline Near Detector, SBND, will characterize the neutrino
beam flux, and have large statistics for neutrino-argon cross section
measurements, while the far detector, ICARUS, will be sensitive to 
the relevant parameter space.
MicroBooNE will continue operating as part of the SBN program, and continue to
deliver valuable information for the design of future
LArTPC experiments, including both the detectors in the SBN program
and the detectors in the DUNE experiment.



\begin{thebibliography}{99}

\bibitem{uboone_tdr}
The MicroBooNE Collaboration, The MicroBooNE Technical Design Report (2012), 
http://www-microboone.fnal.gov/publications.

\bibitem{mboone_lee}
The MiniBooNE Collaboration, Unexplained Excess of Electronlike Events from a 1-GeV 
Neutrino Beam, PRL {\bf 102}, 101802 (2009).

\bibitem{DUNE}
R. Acciari {\it et al.}, Long-Baseline Neutrino Facility (LBNF) and Deep Underground 
Neutrino Experiment (DUNE), arXiv:1512.06148 (2015).

\bibitem{uboone_detector}
R. Acciari {\it et al.}, Design and construction of the MicroBooNE detector,
JINST {\bf 12} (2017).

\bibitem{uboone_stability}
The MicroBooNE collaboration, MicroBooNE Detector Stability,
MICROBOONE-NOTE-1013-PUB (2016).

\bibitem{uboone_noise}
The MicroBooNE collaboration, Noise Characterization and Filtering in the MicroBooNE TPC,
MICROBOONE-NOTE-1016-PUB (2016).

\bibitem{uboone_pandora}
The MicroBooNE collaboration, The Pandora Multi-algorithm Approach to Automated 
Pattern Recognition in LArTPC Detectors, MICROBOONE-NOTE-1015-PUB (2016).

\bibitem{uboone_Michel}
R. Acciari {\it et al.}, Michel Electron Reconstruction Using Cosmic-Ray Data 
from the MicroBooNE LArTPC, arXiv:1704.02927 (2017).

\bibitem{uboone_CCincl}
The MicroBooNE collaboration, Selection and kinematic properties of $\nu_{\mu}$ 
charged-current inclusive events in 5E19 POT of MicroBooNE data,
 MICROBOONE-NOTE-1010-PUB (2016).

\bibitem{uboone_CCpi0}
The MicroBooNE collaboration, Demonstration of 3D Shower Reconstruction on MicroBooNE 
Data, MICROBOONE-NOTE-1012-PUB (2016).

\bibitem{pdg_nuxsec}
G.P. Zeller, Neutrino Cross Section Measurements,
Chin. Phys. C {\bf 40}, 100001 (2016).

\bibitem{SBN}
R. Acciari {\it et al.}, A Proposal for a Three Detector Short-Baseline Neutrino 
Oscillation Program in the Fermilab Booster Neutrino Beam, arXiv:1503.01520 (2015).

\end{thebibliography}
\end{document}